\documentclass[twocolumn,twoside,prd,floatfix,letterpaper]{revtex4}

\usepackage{ifpdf}
\ifpdf 
  \usepackage[pdftex]{graphicx} 
\else 
  \usepackage{graphicx} 
\fi
\usepackage{amsmath}
\usepackage{amssymb}
\usepackage{fancyhdr}
\usepackage{color}
\usepackage{rotating}
\usepackage[colorlinks,hyperindex]{hyperref}

\def \lastDataDate {September 30, 2020}  
\def \suspiciousReturnFigure {Figure~\ref{fig:SuspiciousReturns}}
\def \citeKnuteson {\cite{knuteson2019celebrating,knuteson2018wealth,knuteson2016information}}
\def \citeAllOtherOvernightIntradayLiterature {\cite{cooper2008return,lachance2015night,qiao2020overnight,kelly2011returns,berkman2012paying,branch2012overnight,lou2019tug,bogousslavsky2019cross,lachance2020etfs}}

\begin{document} 

\title{Strikingly Suspicious Overnight and Intraday Returns}
\author{Bruce Knuteson}
\noaffiliation

\begin{abstract}
The world's stock markets display a strikingly suspicious pattern of overnight and intraday returns.  Overnight returns to major stock market indices over the past few decades have been wildly positive, while intraday returns have been disturbingly negative.  The cause of these astonishingly consistent return patterns is unknown.  We highlight the features of these extraordinary patterns that have hindered the construction of any plausible innocuous explanation.  We then use those same features to deduce the only plausible explanation so far advanced for these strikingly suspicious returns.
\end{abstract}

\maketitle

\section{Overnight and Intraday Returns\label{sec:OvernightIntraday}}

Over the past few decades, the world's stock markets have displayed a stunning pattern of overnight and intraday returns.  This stunning pattern is not widely known.  There is no consensus as to the cause.

We have previously provided the only plausible explanation so far advanced for this remarkable pattern of overnight and intraday returns~\cite{knuteson2019celebrating,knuteson2018wealth,knuteson2016information}.  Our previous articles emphasized our explanation, taking for granted our readers' ability to understand that suspicious return patterns in financial markets indicate a problem.  This article describes the extraordinary pattern of overnight and intraday returns in the world's stock markets (Section~\ref{sec:OvernightIntraday}), explains why the most popular innocuous explanations are wrong (Section~\ref{sec:StrikinglySuspicious}), constructs a plausible explanation (Section~\ref{sec:ObviousExplanation}), and concludes that these strikingly suspicious return patterns indicate a serious and urgent problem (Section~\ref{sec:ObviousProblem}).

\suspiciousReturnFigure\ shows overnight and intraday returns to twenty-one major stock market indices around the world.  The intraday (green) curves cumulate returns from market open to market close.  The overnight (blue) curves cumulate returns from market close to the next day's market open.  Over the past two decades, for example, Canada's TSX 60 (top right plot in \suspiciousReturnFigure) has had a whopping +1,062\% return overnight compared to a wrist-slitting intraday return of -67\%.  The rest of the world in \suspiciousReturnFigure\ displays the same general pattern -- large positive overnight returns and large negative intraday returns -- in varying degrees of outrageousness.  The exception is China~\cite{qiao2020overnight,knuteson2019celebrating}.

\suspiciousReturnFigure\ is so egregious you may suspect a mistake, but no mistake has been made.  These shocking return patterns have been noted in the literature for a decade~\citeAllOtherOvernightIntradayLiterature.  You can reproduce them yourself using data publicly available from Yahoo!~Finance and the code at Ref.~\cite{thisArticleWebpage}.  These bewildering patterns are present in indices and in the ETFs that track them.  These amazing patterns are robust to using the price shortly after market open rather than the official open price, to using the price shortly before market close rather than the official close price, and to using data from different data providers.  These and other robustness checks are described in the literature~\citeAllOtherOvernightIntradayLiterature.  The overnight and intraday return patterns in \suspiciousReturnFigure\ are very robust.

\begin{figure*}[p]
\includegraphics[width=7in]{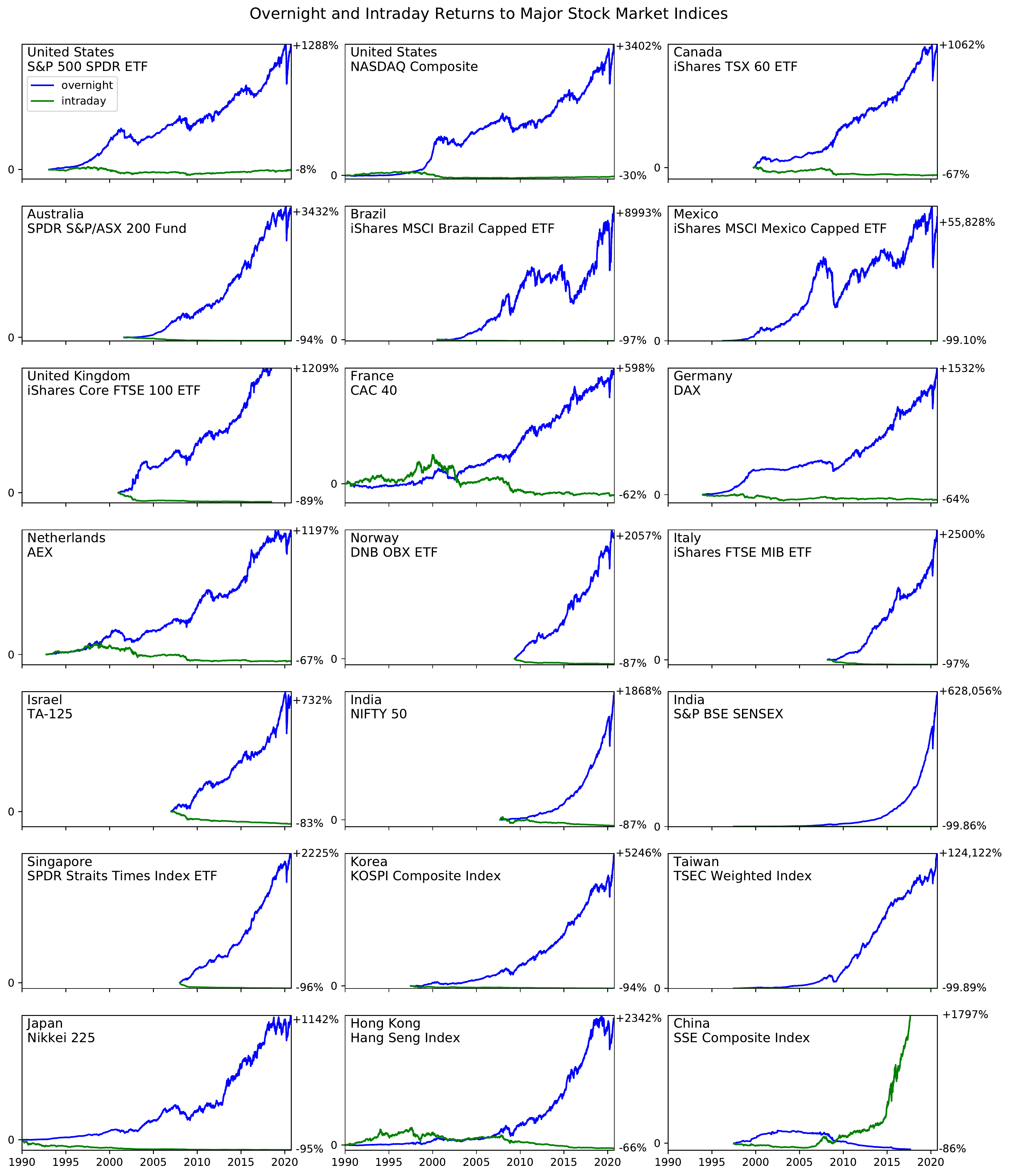}
\caption[Suspicious index returns]{\label{fig:SuspiciousReturns}Cumulative overnight (blue curve) and intraday (green curve) returns to twenty-one major stock market indices over three decades.  Each overnight (blue) curve cumulates returns from market close to the next day's market open.  Each intraday (green) curve cumulates returns from market open to market close.  The horizontal axis of each plot extends from January 1, 1990 to \lastDataDate.  The (linear) vertical scale in each plot extends from a return of -100\% (bottom of plot) through 0 (explicitly marked, at left) to the largest cumulative overnight return achieved (top of plot).  On each plot, the cumulative overnight and intraday returns on \lastDataDate\ (or the last date available) are explicitly marked, at right.  The code used to make this figure is available at Ref.~\cite{thisArticleWebpage}.  Data are publicly available from Yahoo!~Finance~\cite{yahooFinance}.}
\end{figure*}

\section{Strikingly Suspicious\label{sec:StrikinglySuspicious}}

\suspiciousReturnFigure\ is strikingly suspicious.  No variation of ``returns are due to the bearing of risk'' can explain such large negative intraday returns.  The uncoordinated actions of millions of individual traders should not produce such strikingly consistent return patterns.  In this section we consider a few popular innocuous explanations for \suspiciousReturnFigure\ and explain why each is wrong.

Consider the possibility that quarterly company earnings announcements in the United States (outside regular trading hours) are responsible for the top row of \suspiciousReturnFigure~\cite{mccrum2018SomeoneIsWrong}.  You can falsify this attempted explanation by removing the returns around earnings announcements and noting that this does not meaningfully change the plots (as in Section~4.1 of Ref.~\cite{cooper2008return}).  Alternatively, you can simply note that company earnings announcements cannot explain the consistently negative intraday returns in \suspiciousReturnFigure.

Consider the possibility that the spectacular patterns in \suspiciousReturnFigure\ arise because ``almost all price discovery happens overnight.''  You can falsify this attempted explanation by noting that most price movement actually happens intraday~\cite{lachance2015night}~\footnote{If you make a histogram of the overnight returns and the intraday returns for any of the indices in \suspiciousReturnFigure, you will see that the distribution of intraday returns is wider than the distribution of overnight returns.}, not overnight.  Alternatively, you can simply note that ``almost all price discovery happens overnight'' does not explain the consistently negative intraday returns in \suspiciousReturnFigure. 

Consider the possibility that \suspiciousReturnFigure\ arises because of overnight capital costs~\cite{bogousslavsky2019cross}.  You can falsify this attempted explanation for each plot in \suspiciousReturnFigure\ by explicitly including the relevant overnight capital cost and noting that this inclusion is not enough to materially change the picture.   Alternatively, you can simply note that overnight capital costs cannot explain the consistently negative intraday returns shown in \suspiciousReturnFigure.

Consider the possibility that millions of individual ``retail traders'' have systematically different preferences from ``institutional traders,'' that the trades of retail traders are relatively more important earlier in the day than later in the day, and that this persistent difference in preferences is responsible for \suspiciousReturnFigure~\cite{lou2019tug}.  Anyone who has ever interacted with people can falsify this attempt by noting that the patterns in \suspiciousReturnFigure\ are far more consistent than the preferences of millions of individual retail traders over three decades.

Consider the possibility that \suspiciousReturnFigure\ is due to ``ETF arbitrageurs.''  You can falsify this attempted explanation by noting that arbitrageurs do not profit by consistently buying high (as would be necessary to produce the blue curve) and then selling low (as would be necessary to produce the green curve).

Consider the argument that since the suspicious return pattern in United States indices stopped in 2008 (a fact more apparent when the first two plots in \suspiciousReturnFigure\ are plotted with a logarithmic vertical scale~\footnote{Plotting the two United States indices in \suspiciousReturnFigure\ with a logarithmic vertical scale shows two distinct periods.  Before 2008, overnight and intraday returns to the S\&P 500 index and NASDAQ Composite index diverged like those of other indices in \suspiciousReturnFigure.  Since 2008 there has been no such divergence.}), the cause of the suspicious returns no longer matters.  You can point out that the suspicious return pattern has continued in all other indices in \suspiciousReturnFigure\ and in many individual stocks in the United States~\footnote{Apple, Facebook, Google, and Tesla are examples of individual stocks in the United States that exhibit a striking divergence in overnight and intraday returns since 2008.}, so the cause still matters.  You can also point out that this argument does not explain \suspiciousReturnFigure.

Consider finally the assertion that whatever caused \suspiciousReturnFigure\ has not moved stock prices from their ``fundamental value,'' due to efficient markets and all that.  You can point out that current stock prices are impressively high and the difference in total market capitalization between the blue and green curves in \suspiciousReturnFigure\ is roughly fifty trillion dollars.  You can also point out that efficient markets do not look like \suspiciousReturnFigure.

The same logic can be straightforwardly applied to other attempted explanations.  Any attempt that does not explain the large negative intraday returns in \suspiciousReturnFigure\ can be discarded.  Any attempt invoking the actions of millions of individual traders must convincingly reconcile the impressive consistency of \suspiciousReturnFigure\ with the stupefying inconsistency of millions of individual traders over thirty years~\footnote{A preference for numerical tables over plots (which hides the striking consistency so obvious in \suspiciousReturnFigure) and a tendency to view markets as composed of a large number of small participants (when in fact the trading of a small number of large participants dominates the behavior of most markets in the real world) are among the biases that have led economists to unproductively seek explanations for \suspiciousReturnFigure\ involving large numbers of participants.}.  Any attempt invoking the actions of a few traders must explain how consistently buying high and selling low ends up being profitable.

No plausible innocuous explanation for \suspiciousReturnFigure\ has yet been proposed.

\section{Obvious Explanation\label{sec:ObviousExplanation}}

Now that we understand why certain explanations are not plausible, let us reason our way to one that is.

The obvious, mechanical explanation of \suspiciousReturnFigure\ is somebody trading in a way that pushes prices up before or at market open (thus causing the blue curve) and then trading in a way that pushes prices down between market open and market close (thus causing the green curve).  The striking consistency of these plots points to the actions of a few quantitative trading firms rather than the uncoordinated, manual trading of millions of people.  Computers are consistent.  People are not.

Trading in a way that pushes prices up overnight and down intraday is expensive, but the daily expansion and contraction of a sufficiently large portfolio can create mark-to-market gains exceeding this cost~\citeKnuteson.  The resulting risk-return profile is unattractive if the firm is long the market but acceptable if the firm is market neutral~\footnote{The ideal thing to manipulate is something with low volatility that other people cannot easily see.  A single stock has an annual volatility of $\approx$30\% and is easily tracked.  A slowly-time-varying market-neutral portfolio (perhaps also hedged against other major risk factors) has low volatility and is something other people cannot easily see.  As a bonus, there are plenty of ways to get a portfolio optimizer trading thousands of stocks to fall into the appropriate trading pattern -- systematically expanding your portfolio early in the day and then contracting it later in the day -- while remaining willfully blind as to how you are actually making money.}.  As discussed in Ref.~\cite{knuteson2018wealth}, the systematic expansion and contraction of a portfolio that is generally market neutral can leave residual patterns in the overall market like those appearing in \suspiciousReturnFigure.

The obvious suspects are therefore large, market-neutral quant firms that have been around since the early 1990s, trading in volumes large enough to create the remarkable patterns in \suspiciousReturnFigure.  You can read all of this directly off of \suspiciousReturnFigure, and you do not need to be Sherlock Holmes to do it.

This obvious explanation (described in more detail in Refs.~\citeKnuteson) is plausible.  It naturally handles the striking consistency of \suspiciousReturnFigure\ and its large negative intraday returns.  It naturally handles facts you might think would be damning, including how these extraordinary return patterns have not been arbitraged away and why nobody seems to be complaining about them~\cite{knuteson2018wealth}.  It naturally handles China as the exception~\cite{knuteson2019celebrating}.  It naturally handles other related observations~\footnote{The disappearance of the overnight/intraday divergence in United States indices shortly after Ref.~\cite{cooper2008return} called attention to it in 2008 is consistent with the quant firm(s) of Section~\ref{sec:ObviousExplanation} moving some of their morning portfolio expansion from before and at market open to after market open to better hide their footprints, as we recommend in Ref.~\cite{knuteson2019celebrating}.  The continued divergence in the rest of the world is what you might expect from quant firms headquartered in the United States with a habit of applying their domestically focused research to the rest of the world as an afterthought.}~\footnote{The natural explanation for the returns to many trading strategies (e.g., momentum, size, value) displaying an overnight/intraday split~\cite{lou2019tug} is the daily portfolio expansion and contraction described in Section~\ref{sec:ObviousExplanation}, with one or more of the quant firms employing these trading strategies as forecasts or in their risk models.}.  It is falsifiable.  It is easy (for regulators) to check~\footnote{Only a few large, market-neutral quant firms have been around since the early 1990s.  (Including firms that started later or have since closed adds another half dozen or so.)  Systematically expanding your existing portfolio at the start of the day and contracting it later in the day is a distinctive trading pattern.  Quants keep their data, write internal research reports, and communicate by email.  If any large, long-lived quant firm has traded in the manner we describe~\citeKnuteson, somebody knows about it, and there is probably enough documentation to figure out who knew what when.  If no existing, large, long-lived quant firm has traded in this manner, a fairly straightforward analysis of the trading of the most obvious suspects by regulators should be sufficient to falsify this explanation.}.  It is the only plausible explanation so far advanced for \suspiciousReturnFigure.  It is, {\it{ipso facto}}, the leading candidate explanation.  It is everything but innocuous.

\section{Obvious Problem\label{sec:ObviousProblem}}

The world's stock markets display a strikingly suspicious pattern of overnight and intraday returns (Section~\ref{sec:OvernightIntraday}).  No variation of ``returns are due to the bearing of risk'' can explain the large negative intraday returns in \suspiciousReturnFigure.  These plots are too consistent to have been caused by the uncoordinated actions of millions of individual traders.  No plausible innocuous explanation for these extraordinary return patterns has been proposed.  The obvious explanation (Section~\ref{sec:ObviousExplanation}) is obviously problematic.

I seem to be the only person insistently pointing \suspiciousReturnFigure\ out as a problem~\citeKnuteson.  You might conclude from the silence of others that everything is fine.  Fortunately, you can think for yourself, and at some level this issue is not hard.  You understand the difference between large positive numbers and large negative numbers.  You understand the difference between blue lines that go up and green lines that go down.  You understand that strikingly suspicious return patterns in financial markets should be viewed as a problem until definitively shown otherwise, not the other way around.  You know enough (Section~\ref{sec:StrikinglySuspicious}) to evaluate candidate explanations for these strikingly suspicious return patterns for yourself.   Because you can think for yourself, you understand the silence of others does not mean everything is fine.  The world's stock markets display a strikingly suspicious pattern of overnight and intraday returns (\suspiciousReturnFigure).  I appear to be the only person persistently trying to alert you to them.  Everything is pretty far from fine.

\bibliography{strikingly_suspicious}

\end{document}